\documentclass[twocolumn,showpacs,preprintnumbers,amsmath,amssymb]{revtex4}
\newcommand{\be}{\begin{equation}}
\newcommand{\ee}{\end{equation}}

\newcommand{\bwt}{\begin{widetext}}
\newcommand{\ewt}{\end{widetext}}

\newcommand{\p}{\partial}

\usepackage{epsf}
\usepackage{graphics}

\begin{document}

\preprint{IMSc/2003/07/23}

\vskip.5cm
\title{\huge{\bf Optical Activity From Extra Dimension.}}

\author{ Avijit K. Ganguly}
\email{avijit@imsc.res.in} 
\author{R. Parthasarathy}
\email{sarathy@imsc.res.in} 

\affiliation{ Institute Of Mathematical Sciences, Taramani, Chennai 600113,
India.}

\vskip0.2cm

\vglue 0.3truecm
\begin{abstract}
\begin{center}
{\bf{Abstract }}
\end{center}
\noindent
Optical activity, like Faraday effect, is a rotation of the plane of 
polarization of propagating light in a medium and can be attributed 
to different sources with distinct signatures. In this note we discuss 
the effect of optical activity {\it{in vacuum}} due to Kaluza-Klein 
scalar field $\phi$, in the presence of an external electro-magnetic field.
The astrophysical implication of this effect is indicated. We also point out
 the possibility of observing the same in laboratory conditions.
\\
\end{abstract}

\maketitle

Higher dimensional gravity theories have been
considered as possible avenues to unify the basic forces of nature. 
5-d Kaluza- Klein (KK) \cite{kk} theory unifies gravity and electromagnetism
 and extensions of this have been investigated in \cite{thoms}. The 
extra dimensions have been assumed to be small, typically of the
order of Planck length and so the KK modes are
highly massive. The successful string theories, constructed in 10 or
26 dimension, encompass the higher dimensions. Nevertheless, the 
extra dimensions will not be directly observable in experiments. The success 
of string theories gave encouragement to search for indirect methods to detect
the extra dimensions.  Possible effects of the extra dimensions considered 
as bulk in the Standard Model have been suggested by Arkani-Hamed, Dimopoulos 
and Dvali \cite{ADD}. Various predictions in the standard model
 as signatures of Higher dimensions have been studied \cite{AADD}. For example, in order to
resolve the hierarchy problem, large extra dimensions of the order of sub-millimeter have been
proposed. Recently Matsuda and Seki \cite{MAT} studied the Casimir energy of the extra dimensions
and related this to the 4-dimensional cosmological constant ${\Lambda}_c$. Using the WMAP data on
${\Lambda}_c$, they estimated the size of the extra dimensions to be of the order of
sub-millimeter region.  
In all these
 studies a specific model has been used like
Standard Model and the deviations due to the extra dimensions are small,
in a large background. It is worth while to investigate a situation which 
is largely due to the effects of the extra dimensions. It is the purpose here 
to consider the simplest higher dimensional theory, viz, 5-dimensional KK 
theory. The '55' component of the 5-dimensional metric is taken to be dependent
 on the  physical dimension (t, x,y,z). Then the standard procedure of taking 
$X^5$ as a circle of radius R, yields a coupling of the type $\phi 
F_{\mu\nu}F^{\mu\nu}$ where $F^{\mu\nu}$ is the electromagnetic field strength
and $\phi$ is the relic of the '55' component of the 5-dimensional metric. 
 The effect of the term can be studied by writing $F_{\mu\nu} \to F_{\mu\nu} +
f_{\mu\nu}$, where $F_{\mu\nu}$ is a background EM-field (which is taken to be 
pure magnetic) and $ f_{\mu\nu} $ represent a propagating  laser beam. Then 
the above coupling induces a change in the polarization plane of the laser 
beam even in homogeneous medium or vacuum and also a change in the velocity of 
light in preferred directions. {\it
The change in the polarization plane
is purely due to the above coupling and so constitute an
indication of the effect of extra dimensions}\footnote{Which
we feel could be distinguished from the same coming from other
sources for example axions or neutrinos or Faraday effect.}\\
%
%
%

\noindent
In this note we closely follow the notations of
\cite{chodos}. The action for higher $(4+n)$ dimensional gravity  
can be written in the
following form,
\begin{eqnarray}
S = \frac{1}{16 \pi G_{4+n}}\int d^{4+n}x \sqrt{|\gamma|}R, 
\label{action}
\end{eqnarray}
where $\gamma$ in $4+n$ dimensional metric, $R$ is the 
scalar curvature computed from $\gamma$ and  
 $G_{4+n}$ is the $(4+n)$-dimensional Newton's constant.  Explicit
form of the metric for a 5-dimensional theory is given below. 
\begin{displaymath}
\mathbf{\gamma_{mn} = 
\phi^{-\frac{1}{3}}\left[ \begin{array}{cc}
 g_{\mu\nu}+ \kappa^2 \phi A_{\mu} A_{\nu}\ & \kappa \phi A_{\mu} \\
\kappa \phi A_{\nu} 
  & \phi \\
\end{array} \right]}
\end{displaymath}
with $g_{\mu\nu}$ as the usual 4-dim metric, the vector fields
$A_{\mu}$ are the electro magnetic potentials and $\phi$ is the 
scalar field from ${\gamma}^{55}$. It should 
be noted that none of the fields that appear in the above 
expression  are  functions  of the  extra  compact coordinate.   

\vspace{0.5cm}

Given the form of the metric, one can compute the Ricchi scalar
$R$ and write  down the 4 dimensional action given by (after integration 
the extra dimension):
\begin{eqnarray}
S = \frac{1}{16 \pi G_{4}}\int d^{4}x \sqrt{- g}
\left[R^{(4)} + \frac{1}{4} \phi F_{\mu\nu}F^{\mu\nu} + \frac{1}{6}
\frac{\p_{\mu}\phi \p^{\mu}\phi}{\phi^2}\right].
\label{action}
\end{eqnarray}

In what follows, we consider a flat Minkowski space for the 4-dimensional space-time.   
Subsequently the scalar field
$\phi$ is scaled by a dimensionful constant $M^2$ as, 
\begin{eqnarray}
\rm{log}[\phi M^2] = \psi. 
\label{scaling}
\end{eqnarray}

\vspace{0.5cm}

Then the  classical action reduces to:
\begin{eqnarray}
S = \frac{1}{16 \pi G_{4}}\int d^{4}x
\left[ \frac{1}{4 M^2} e^{\psi} F_{\mu\nu}F^{\mu\nu} +\frac{1}{6}
\p_{\mu}\psi \p^{\mu}\psi \right].
\label{reduced-action}
\end{eqnarray}
Here $\psi$ bears the effect of the relic of fifth dimension.
Next if one re-scales the field $\psi$ as 
$\psi \rightarrow \frac{\sqrt{6}}{M} \psi$ and define  
$\lambda =\frac{\sqrt{6}}{4M}$ with $M^2=\frac{1}{16\pi G_{4}}$, then 
expanding in powers of $\lambda$,
from eqn (\ref{reduced-action})we have,  the Lagrangian:
%
%
%
%
%
%
%

%
\begin{eqnarray}
{\cal{L}}= \frac{1}{4} F_{\mu\nu}F^{\mu\nu} +
\lambda \psi F_{\mu\nu}F^{\mu\nu}  +
\p_{\mu}\psi \p^{\mu}\psi,
\label{reduced-action1}
\end{eqnarray}
which is of the same form as obtained  in Miani, Petronzio and Zavatini 
\cite{miani}.
%
%
%
%
%
 In the presence of an external background electro magnetic field, $F^{\mu\nu}$, 
eqn. (\ref{reduced-action1}) can be expressed in the following form:
\bwt
\begin{eqnarray}
{\cal L} = \frac{1}{4}F_{\mu\nu}F^{\mu\nu}+ \frac{1}{2} f_{\mu\nu}F^{\mu\nu}
+ \frac{1}{4}f_{\mu\nu}f^{\mu\nu}  
+ \lambda \psi \left[ F_{\mu\nu} F^{\mu\nu}+ 
2 f_{\mu\nu}F^{\mu\nu} + f_{\mu\nu}f^{\mu\nu}\right] +
\p_{\mu}\psi\p^{\mu}\psi .
\label{act-ext}
\end{eqnarray}
\ewt
In eqn. (\ref{act-ext}) 
$f^{\mu\nu}$ represents the dynamical
field given by, $f^{\mu\nu}=\partial^{\mu}A^{\nu}- \partial^{\nu}A^{\mu}$,
where $A^{\mu}$ corresponds to the fluctuation. The equations of motion obtained
from eqn. (\ref{act-ext}) neglecting $f^{\mu\nu}f_{\mu\nu}\psi$ are,  

\begin{eqnarray}
\p_{\mu} \p^{\mu} \psi = 2\lambda f_{\mu\nu}F^{\mu\nu} \nonumber \\
\left[\p_{\alpha} \p^{\alpha} g^{\nu}_{\mu} - \p_{\mu} \p^{\nu}\right]A^{\mu}
= -8 \lambda \p_{\mu}\psi F^{\mu\nu}.
\label{eom}
\end{eqnarray}
\vspace{0.5cm}

\noindent
When the external magnetic field is constant, one can write down 
eqn.(\ref{eom}) in momentum space:
\begin{eqnarray}
 k^{2} \psi &=& -4i\lambda k_{\mu}A_{\nu}F^{\mu\nu} \nonumber \\
\left[k^{2} g^{\nu}_{\mu} - k_{\mu} k^{\nu}\right]A^{\mu}
&=& -8 i \lambda k_{\mu}\psi F^{\mu\nu}.
\label{eom-m}
\end{eqnarray}
It should however be noted that till now no explicit gauge choice 
has been made. However we will discuss this issue later. As a next step
we eliminate the field $\psi$ from equations, (\ref{eom-m}),to 
arrive at,
\begin{eqnarray}
\!\! \left[k^{2} g^{\nu}_{\mu} - k_{\mu} k^{\nu}\right]A^{\mu} \!\!\!
&=&\!\!\! -8i\lambda k_{\mu} \frac{\left[-4i\lambda k_{\alpha}A_{\beta}
F^{\alpha\beta} \right]}{k^2} F^{\mu\nu}.
\label{eom-m}
\end{eqnarray}

\vspace{0.5cm}

For an external magnetic field in the z direction, the only nonzero component
 of the field strength tensor is $F^{12}$. Therefore one can introduce the 
following four-vectors, $B_{\lambda}= (0,0,0,1)$ and $u_{\rho}=(1,0,0,0)$  
, to write down the external field strength tensor,
$F^{\mu\nu}$ as $F^{\mu\nu} = |{\cal B}|\epsilon^{\mu\nu\lambda\rho} 
B_{\lambda}u_{\rho}$.\\

%
%
\noindent
Discussion of optical activity is usually described in a plane, formed by
 the polarization vectors such that they are
orthogonal to the propagating direction $k^{\mu}$ of the field. 
Using the available vectors, we define them in the following way
\cite{faraday}, 
\begin{eqnarray}
e_{\nu,1} = N \epsilon_{\mu\nu\lambda\sigma}k^{\nu}B^{\lambda}u^{\sigma}
\label{e1} \\
e_{\nu,2} =N' \epsilon_{\mu\nu\lambda\sigma}k^{\nu}B^{\lambda}e^{\sigma}_{1}
\label{e2}
\end{eqnarray}
%
%
such that, the set of three vectors, $k^{\mu}$, $e^{\mu}_{1}$,
$e^{\mu}_{2}$ form a orthogonal combination. The normalizations are $N=i/{|k_{\perp}|}$ and
$N'=1/{\sqrt{(k_0^2-k_{\perp}^2)}}$ where $k_{\perp}^2=k_x^2+k_y^2$.    
We construct another 4-vector, 
\begin{eqnarray}
\hat{B}^{\rho} =\left[B^{\rho} - 
\frac{k.B}{k^2} k^{\rho} \right]
\label{bhat}
\end{eqnarray}
and verify that,
$$ e_1.e_2= e_1.k=e_2.k=k.\hat{B}=\hat{B}.e_1= \hat{B}.e_2 =0.$$
Thus we have  a system of four vectors which are orthogonal
to each other. With these,  it is possible to expand
the four vector describing the dynamical field, $A^{\mu}$ as,
\begin{eqnarray}
A^{\alpha} = A_1 e^{\alpha}_1 + A_2 e^{\alpha}_2 + A_3 \hat{B}^{\alpha}
+ A_{||} k^{\alpha}.
\label{dyna}
\end{eqnarray}
In eqn.(\ref{dyna}), $ A_i$'s ( i=1,..3) are scalars depending on  
the four vectors $k^{\mu}$ and $F^{\mu\nu}$. 
Substituting (\ref{dyna}) in 
eqn.(\ref{eom-m}), we arrive at,
\bwt
\begin{eqnarray}
k^4 (A_1 e^{\nu}_1 + A_2 e^{\nu}_2 + A_3 \hat{B}^{\nu})=
-32\lambda^2 k_{\mu}F^{\mu\nu} \left[
k_{\alpha}\left(e_{\beta 1}+ A_2e_{\beta 2}\right) \epsilon^{\alpha\beta\lambda\rho}
B_{\lambda}u_{\rho}
\right]
\label{eom-f}
\end{eqnarray}
\ewt
\vspace{0.5cm}

It should be noted further that, the component $A_{||}$ associated with
the gauge degree of freedom has finally dropped out. So without loss
of generality, one can choose $A_{||} =0$, so that the Lorentz
gauge condition $k.A=0$ is satisfied. Further, as we contract
eqn. (\ref{eom-f}) with $\hat{B}_{\nu}$, it gives,
\begin{eqnarray}
k^4  A_3 = 0.
\label{gl}
\end{eqnarray}
Since we are working with an 
interacting theory, $k^2$ is nonzero and  therefore
$A_3 = 0$.
Hence one is left with two degrees of freedom, as
should be the case for any physical photon in absence of a medium.\\

\noindent
Next one can obtain the equations of motion for the transverse pieces
by contracting eqn. (\ref{eom-f}) by $e_{\nu 1}$ and $e_{\nu 2}$.
Using the form of external field strength tensor, $F^{\mu\nu}$ expressed
in terms of the four vectors $B^{\rho}$ and $u^{\alpha}$ as has been
 given before, one finds them to be:
\bwt
\begin{eqnarray}
k^4 A_1 (e_1)^2 = - 32\left(\lambda {\cal B}\right)^2 \epsilon^{\mu\nu\delta\rho}k_{\mu}
e_{\nu 1}B_{\delta}u_{\rho} \left[ \epsilon^{\alpha\beta\lambda\sigma}
k_{\alpha}e_{\beta 1} B_{\lambda} u_{\sigma} A_1 +
\epsilon^{\alpha\beta\lambda\sigma}
k_{\alpha}e_{\beta 2} B_{\lambda} u_{\sigma} A_2 \right] \\
k^4 A_2 (e_2)^2 = - 32\left(\lambda {\cal B}\right)^2 \epsilon^{\mu\nu\delta\rho}k_{\mu}
e_{\nu 2}B_{\delta}u_{\rho} \left[ \epsilon^{\alpha\beta\lambda\sigma}
k_{\alpha}e_{\beta 1} B_{\lambda} u_{\sigma} A_1 +
\epsilon^{\alpha\beta\lambda\sigma}
k_{\alpha}e_{\beta 2} B_{\lambda} u_{\sigma} A_2 \right] 
\label{coupled}
\end{eqnarray}
\ewt
Denoting  
\begin{eqnarray}
Q_1= \sqrt{32 \left(\lambda {\cal B}\right)^2} \epsilon^{\mu\nu\delta\rho}k_{\mu}e_{\nu 1}B_{\delta}
u_{\rho} \\
Q_2= \sqrt{32  \left(\lambda{\cal B}\right)^2}\epsilon^{\mu\nu\delta\rho}k_{\mu}e_{\nu 2}B_{\delta}
u_{\rho} 
\label{Q}
\end{eqnarray}
the coupled equations (\ref{coupled}) can now be 
written in matrix notation as,
\begin{eqnarray}
\left[ \begin{array}{cc}
 k^4 + Q^2_1 & Q_1 Q_2 \\
  Q_1 Q_2
  &   k^4 +Q^2_2  \\
\end{array} \right]
\left[ \begin{array}{c}
A_1 \\
A_2
\end{array} \right]=0.
\end{eqnarray}
From the secular determinant one obtains the following relation, 

\begin{eqnarray}
 k^4 +(Q^2_1 + Q^2_2)=0. 
\label{Fn1}
\end{eqnarray}
In order to evaluate the dispersion relation from eqn. (\ref{Fn1}), one  
needs to find out the quantities $Q_1$ and $Q_2$.  
 By making use of  eqns. (\ref{e1}) and (\ref{e2}) in (\ref{Q}), one 
readily finds out that,
\begin{eqnarray}
 Q_1 = i\sqrt{ \left(32\lambda^2{\cal B}^2 \right)} |k_{\perp}|
\mbox{~~and~~} Q_2=0.  
\label{Q1Q2}
\end{eqnarray}
The dispersion relations for the photons following from, eqn. (\ref{Fn1}) are 
given by,
\begin{eqnarray}
k^2_0 - \vec{k}^2 = \sqrt{32}\lambda |{\cal B}| |k_{\perp}| \nonumber \\ 
\label{disp1}
k^2_0 - \vec{k}^2 = - \sqrt{32}\lambda |{\cal B}| |k_{\perp}|   
\label{disp2}
\end{eqnarray}
As can be seen from eqns. (\ref{disp1}) and (\ref{disp2}), the effect of
 the KK mode and the magnetic field are actually responsible for the deviation
from the normal dispersion relation in vacuum. In fact a closer look at the 
same equations also reveal that if the propagation direction of the 
photon is along the magnetic field, one essentially recovers the 
vacuum dispersion relation. So there is no birefringence effect or in other
words no optical activity in this case. On the other hand for any finite value of 
$|k_{\perp}|$, when the propagation direction of the light is not along the magnetic field, the
coupling induces optical activity even in vacuum.   
Consequently the velocities $v$ defined as $\frac{k_0}{\vec{k}}$ of the two propagating modes are,
\begin{eqnarray}
v^2_{(1)} = 1+ \sqrt{32}\lambda |{\cal B}|\frac{ |k_{\perp}|}{\vec{k}^2} 
\nonumber \\ 
\label{disp1}
v^2_{(2)} = 1 - \sqrt{32}\lambda |{\cal B}| \frac{|k_{\perp}|}{\vec{k}^2}   
\label{disp2}
\end{eqnarray}
 In the limit when  $ \left(\sqrt{32}\lambda |{\cal B}|\frac{ |k_{\perp}|}{\vec{k}^2} \right)
< 1 $, the velocity difference between the two modes turn out to be,
\begin{eqnarray}
\Delta v =  v_{(1)}-  v_{(2)} =  \sqrt{32}\lambda |{\cal B}| \frac{|k_{\perp}|}{\vec{k}^2}              
\end{eqnarray}
In other words, the presence of external magnetic field makes the space 
anisotropic with reference to the direction of the magnetic field and the  
light propagates with different speeds in the parallel and    
 perpendicular directions to the magnetic field. In the same limit, the phase lag between
 the two polarization modes given by \cite{faraday},
\begin{eqnarray}
\Delta k = \sqrt{32}\lambda |{\cal B}| \frac{|k_{\perp}|}{k_0}.              
\end{eqnarray}

\noindent
Interestingly enough, the time lag suffered by the two polarized
 plane wave fronts coming from a distance $l$ would be given by:
\begin{eqnarray}
\Delta t = l \left( \frac{1}{v_{(1)}}- \frac{1}{ v_{(2)}} \right) 
= - \frac{ \Delta v}{v_{(1)}v_{(2)}} 
\label{timedelay}
\end{eqnarray}
\noindent
The rotation of plane of polarization $\Phi$, in the limit as mentioned 
before, turns out to be:
\begin{eqnarray}
\Delta \Phi= \Delta k l = 
\sqrt{32}\lambda |{\cal B}| \frac{|k_{\perp}|}{k_0}l  .
\label{rotation}
\end{eqnarray}
One can now define a ratio, $\frac{\Delta t}{\Delta \Phi}$, that in the
strict limit of $\lambda \ne 0$, turns out to be:
\begin{eqnarray}
\frac{\Delta t}{\Delta \Phi}= \frac{k_0}{\vec{k}^2} \left( 1+ \frac{1}{2} 
\left(\sqrt{32}\lambda |{\cal B}|\right)^2 \frac{|k^2_{\perp}|}{\vec{k}^4} 
\right) .
\label{ratio}
\end{eqnarray}
Eqn.(\ref{ratio}) is interesting from the point of view that, the differential 
time delay and rotation of plane of polarization can be simultaneously 
used to draw some inference about the effect of the extra dimension.
We believe that this effect can have observable consequences in astrophysical 
situations as well.  A possible
dispersion of EM waves in vacua with velocity difference attributed to quantum gravity effects was explored in \cite{AME}. The same was used to extract a bound on quantum gravity mediated corrections. In this note we 
explore similar velocity differences with two modes of polarization
 in the context of higher dimensional theories.  In view of the
recent  polarization studies of Gamma Ray bursters as reported in \cite{eli}
it is felt that a correlated study of energy dependent velocity difference 
vs polarization of the high energy Gamma Rays from cosmological gamma ray 
bursters can be used to put a bound on the higher dimensional coupling 
constant.  
We also note that one can test the same using some 
laser interferometric experiments in laboratory conditions. However the 
details of these  studies are too involved to be dealt with in this 
note and would be taken up in a subsequent investigation.\\

\begin{acknowledgements} We would like to thank Prof. G. Rajasekaran for many
 valuable discussions and  his 
 interest in this work.  
\end{acknowledgements}


\end{document}